\documentstyle[11pt]{article}

\def\be{\begin{equation}}
\def\en{\end{equation}}
\def\bem{\begin{displaymath}}
\def\enm{\end{displaymath}}
\def\ket#1{| #1 \rangle}
\def\bra#1{\langle #1 |}
\def\BraKet#1#2{\langle #1 | #2 \rangle}

\def\Tr{{\rm Tr}}

\begin{document}
\centerline{\Large\bf Unconditionally Secure Relativistic Quantum 
                      Bit Commitment}
\vskip 3mm
\centerline{S.N.Molotkov and S.S.Nazin} 
\vskip 2mm
\centerline{\sl\small Institute of Solid State Physics of 
                      Russian Academy of Sciences,}
\centerline{\sl\small 142432 Chernogolovka, Moscow District, Russia}
\vskip 3mm

\begin{abstract}
A new relativistic quantum protocol is proposed allowing to implement
the bit commitment scheme. The protocol is based on the idea that
in the relativistic case the field propagation to the region of space
accessible to measurement requires, contrary to the non-relativistic case,
a finite non-zero time which depends on the structure of the particular state 
of the field. In principle, the secret bit can be stored for arbitrarily long 
time with the probability arbitrarily close to unit.
\end{abstract}
\vskip 2mm\noindent
e-mail: molotkov@issp.ac.ru, nazin@issp.ac.ru
\vskip 3mm
\section{Introduction}

The bit commitment scheme includes two distant users A and B and can be
described as follows [1]. At the time moment $t_0$ user A choses one bit 
(0 or 1) and then sends to 
user B only partial information on that bit so that this information
is not sufficient to find out which bit was actually chosen by user A.
At the second (commitment) stage user B can ask A to send him the rest 
information on the bit. The protocol should provide guarantees to user B
that at the second stage user A did not change his mind on the bit chosen
at $t_0$, so that user A cannot cheat.

The classical bit commitment schemes when A and B can only exchange information 
through the classical channel are based on the unproven computational complexity 
of e.g. discrete logarithm problem. The schemes which employ a quantum 
information channel in addition to the classical one are called quantum
bit commitment schemes. In that case information is carried by quantum states.
Earlier various quantum bit commitment protocol have been proposed [2--4].
However, later the impossibility of the ideal non-relativistic
quantum bit commitment protocol was proved since user A can always cheat 
user B by employing the EPR-pairs and delaying his measurement
(actually delaying his choice of the bit) until the second stage of 
the protocol [5,6].

Recently, a bit commitment protocol was proposed which takes into account 
the finite speed of signal (information) propagation [7,8]. 
This relativistic classical protocol is unconditionally secure
(its security is based on the fundamental laws of nature only)
and allows to delay the second stage of the protocol (i.e. to store
the information on the secret bit chosen by user A) for arbitrarily
long time, although it requires that the two parties A and B each have 
a couple of spatially separated sites fully controlled by them 

Below we propose a relativistic quantum protocol allowing
to realize the bit commitment scheme during a finite time interval.
To be more precise, in the indicated protocol the user B cannot 
determine the bit chosen by user A employing the information sent 
to him by user A during a finite time $t_c$ (in other words, 
the probability of 
correct bit identification during arbitrarily large but finite chosen 
in advance time interval $t_c$
does not exceed the probability of correct guessing of that bit, i.e. 1/2, by
arbitrarily small chosen beforehand quantity $\varepsilon$). On the other hand, 
user B can ask user A at any time $t<t_c$ which bit he had actually chosen 
and later at any moment $T>t_c$ he can check which bit was chosen by user A
at $t_0$. It should be noted that actually the finite time $t_c$ is also
present in the classical bit commitment protocols based on computational 
complexity of various problems where it is determined by the computational
resources available for user B and should therefore be selected in such a
way that the identification of the chosen bit from the information disclosed 
by user A is impossible in a time interval shorter than $t_c$.
In our protocol the impossibility of identification of the secret bit
in a time interval shorter than $t_c$ is based on the fundamental laws of 
nature (like in the classical relativistic protocol [7,8]) rather than
the advances in technology available for user B; however, employment of the
quantum states (to be more precise, single-particle quantum field states)
as the information carriers allows to construct a protocol where each user
controls a single site. In addition, in contrast to the non-relativistic 
quantum mechanics, the constraint on the maximum speed of
information transfer allows to develop a secure protocol based on the 
orthogonal states.

The paper is organized as follows. Section 2 deals with
 quantum-mechanical
measurements employed in the protocol; a simple model 
one-dimensional measurement is described in Section 3. The protocol itself
is given in Section 4.

\section{Quantum-mechanical measurements used in the protocol}

All quantum cryptographic protocols actually employ the following two
features of quantum theory. The first one is the no cloning theorem [16], 
i.e. the impossibility
of copying of an arbitrary quantum state which is not known beforehand or,
in other words, the impossibility of the following process:
\bem
\ket{A}\ket{\psi} \rightarrow U(\ket{A}\ket{\psi})=
                 \ket{B_{\psi}}\ket{\psi}\ket{\psi},
\enm
where $\ket{A}$ and $\ket{B_{\psi}}$ are the
apparatus states before and copying act, respectively, and
$U$ is a unitary operator. Such a process is prohibited
by the linearity and unitary nature of quantum evolution.
Actually, even a weaker process of obtaining any information about
one of the two non-orthogonal states without disturbing it is impossible,
i.e. the final states of the apparatus $\ket{A_{\psi_1}}$ 
and $\ket{A_{\psi_2}}$ corresponding to the initial input states
$\ket{\psi_1}$ and $\ket{\psi_2}$, respectively, after the unitary 
evolution $U$, 
\bem
\ket{A}\ket{\psi_1} \rightarrow U(\ket{A}\ket{\psi_1})=
                         \ket{A_{\psi_1}}\ket{\psi_1} ,
\enm
\bem
\ket{A}\ket{\psi_2} \rightarrow U(\ket{A}\ket{\psi_2})=
                         \ket{A_{\psi_2}}\ket{\psi_2} ,
\enm
can only be different, $\ket{A_{\psi_1}} \neq \ket{A_{\psi_2}}$, if 
$\BraKet{\psi_1}{\psi_2} \neq 0$ [10], which means the impossibility 
of reliable distinguishing between non-orthogonal states.
There is no such a restriction for orthogonal states.
That is why almost all cryptographic protocols employ
non-orthogonal states as information carriers, the only 
exception being the protocol suggested in Ref.[11].

In quantum mechanics any two orthogonal states can be reliably distinguished, 
and within the framework of the non-relativistic theory this can be done 
instantly. At the present time there is every ground to believe that
in the relativistic quantum theory the orthogonal quantum states
can also be reliably distinguished without disturbing them. However,
the existence of a finite maximum speed of the information transfer and 
field propagation imposes a restriction on the time required for this process:
two orthogonal states can only be reliably distinguished during a finite 
non-zero time interval whose duration depends on the structure of the states.
The reason is that such measurements are actually non-local, as explained below.
 
Consider first the non-relativistic case. Suppose we are given two orthogonal 
states of a one-dimensional non-relativistic particle
$\ket{\psi_1}$ and $\ket{\psi_2}$,
\be
\BraKet{\psi_1}{\psi_2}=0.
\en
In the momentum realization of the Hilbert space we have
\be
\ket{\psi_{1,2}}=\int_{-\infty}^{\infty}\psi_{1,2}(p)\ket{p}dp.
\en
Let the wavefunctions $\ket{\psi_{1,2}}$ have non-overlapping supports 
\be
\mbox{supp}\mbox{ }\psi_1(p)\bigcap \mbox{supp}\mbox{ }\psi_2(p)=\emptyset.
\en
\bem
\BraKet{\psi_1}{\psi_2}=\int_{-\infty}^{\infty}\psi^{*}_{1}(p)\psi_{2}(p)dp=0.
\enm
The orthogonality of the states is generally a non-local property in the sense
that the scalar product involves the values of the wavefunctions at
all points in the space. If the states are orthogonal in the entire space,
their projections on a particular subspace should not necessarily be orthogonal. 
To illustrate this point, consider the same states in the position realization
of the Hilbert space of states:
\be
\ket{\psi_{1,2}}=\int_{-\infty}^{\infty}\psi_{1,2}(x)\ket{x}dx,\quad
\psi_{1,2}(x)=\int_{-\infty}^{\infty}e^{ipx}\psi_{1,2}(p)dp.
\en
Generally, the orthogonality is only preserved in the entire space
and the same states restricted to a certain subspace should not be
necessarily orthogonal:
\be
\BraKet{\psi_1}{\psi_2}=\int_{\Omega(x)}\psi^{*}_{1}(x)\psi_{2}(x)dx\neq 0,
\quad\mbox{if}\quad \Omega(x)\neq(-\infty,\infty).
\en
Reliable distinguishing of the two states generally requires the access
to the entire region of space where these states are present. Non-relativistic
quantum mechanics allows instantaneous access to the entire coordinate space,
that is it allows instantaneous non-local measurements. To be more precise, 
according to the standard non-relativistic quantum-mechanical measurement
theory the outcome statistics refers to the system state at a particular moment
of time. In the non-relativistic case the information on the outcome of 
a certain measurement involving distant points (e.g. the measuring apparatus 
can ``read off'' information on a quantum state simultaneously in different
spatial points) can be gathered instantly to an observer at a certain point
as long as there are no restrictions on the speed of information transfer.

Unlike the non-relativistic quantum mechanics, there is still no 
consistent measurement theory for the relativistic case. The instantaneous 
non-local measurements in the sense described above for the non-relativistic 
quantum mechanics seem to be impossible in the relativistic quantum theory.
Even if the measuring apparatus performs simultaneous (in a certain reference 
frame) measurement of a quantum state in two different points
${\bf x}_1$ and ${\bf x}_2$, the information on the obtained outcomes
cannot be gathered by a single localized observer in a time interval 
shorter than $t=|{\bf x}_1-{\bf x}_2|/2c$. 
Therefore, non-local measurements require finite time which depends on 
the structure of the measured states.

Thus, in the relativistic quantum field theory gathering information on the
outcome of a non-local measurement requires a finite time. In addition,
if the measuring apparatus has only access to a limited domain $D$ of space,
the measurement performed over a non-stationary state would only provide 
information on that state if its spatial supports has common points with $D$ 
at the time of measurement. If the state support lies outside $D$ at time $t_0$,
the full information on the state employing the measurements restricted to $D$
is only possible in time $t_1\approx L/2c$, where $L$ is the size of the state
support since the state cannot move into the domain $D$ in less than $t_1$
because of the constraint on the maximum field propagation speed.

Before describing the protocol, we shall first discuss the measurements 
it employs. In the rest of the paper we shall deal with the massless particles
(photons). The four-dimensional vector-potential operators have the form [12]
\be 
A^{\pm}_{n}(\hat{x})=\frac{1}{(2\pi)^{3/2}}\int 
e^{\pm i\hat{k}\hat{x}}A^{\pm}_{n}({\bf k}) \frac{d{\bf k}}{\sqrt{2k^0}}=
\frac{1}{(2\pi)^{3/2}}\int e^{\pm i\hat{k}\hat{x}}e^m_n({\bf k})
a_{m}^{\pm}({\bf k})\frac{d{\bf k}}{\sqrt{2k^0}},
\en
where ${\hat k}\hat{x}=k^0x^0-{\bf kx}$. 
The operators $A_{n}^{\pm}(\hat{x})$ 
are supposed to satisfy the Bose commutation relations [12]
\be
[A_{m}^{-}({\hat x}_1),A_{n}^{+}({\hat x}_2)]_-=ig_{mn}D^-_0(\hat{x}_1-\hat{x}_2),
\en
where $g^{nm}$ is the metric tensor ($g^{00}=-g^{11}=-g^{22}=-g^{33}=1$),
$D_0^-$ is the negative-frequency commutator function
for the massless field
\be
D_0^-(\hat{x})=-i\frac{1}{(2\pi)^{3/2}}\int 
d\hat{k}\delta(\hat{k}^2)\theta(-k^0)e^{i\hat{k}\hat{x}}=
\frac{i}{(2\pi)^3}\int \frac{d{\bf k}}{2|{\bf k}|}e^{-ix^0|{\bf k}|+i{\bf xk} },
\en
which is only different from zero and has a singularity on the light cone [12]
\be
D^-_0(\hat{x})=\frac{1}{4\pi}\varepsilon(x^0)\delta(\lambda),
\quad \lambda^2=(x^0)^2-{\bf x}^2.
\en
Further we put $c=1$. The creation and annihilation operators 
$a^{\pm}_{m}({\bf k})$ describe the photons of four types:
two transverse, one longitudinal, and one temporal. The two latter types
are non-physical and are only introduced to preserve the four-dimensional
structure of the vector-potential. The commutation relations are
\bem
[a^{-}_{m}({\bf k}),a^{+}_{n}({\bf k}')]_-=-g^{nm}\delta({\bf k}-{\bf k}'),\quad
({\bf e}^{\alpha}\cdot{\bf e}^{\beta})=\delta_{\alpha,\beta},\quad
(\alpha,\beta=1,2,3),\quad e^{\alpha}_{0}=0,\quad{\bf e}^{3}=\frac{\bf k}{|{\bf k}|}.
\enm
When working with the four types of photons one should employ the indefinite 
metric. For our purposes it is sufficient to deal with a single sort
of photons with a fixed helicity and we shall therefore work in
the subspace of one-particle states equipped with a standard hermitian 
scalar product (for details see Ref.[12]).

The one-photon states corresponding to the bit values 0 and 1, respectively,
$\ket{\psi_1}$ and $\ket{\psi_2}$, can be chosen in the form
\be
\ket{\psi_{1,2}}=\int \psi_{1,2}({\bf k})a^+({\bf k},s)\ket{0}
\frac{d{\bf k}}{\sqrt{2k^0}}=\int \psi_{1,2}({\bf k})\ket{{\bf k},s}
\frac{d{\bf k}}{\sqrt{2k^0}},\quad
\BraKet{{\bf k},s}{{\bf k}',s}=\delta({\bf k}-{\bf k}'),
\en
where $a^+({\bf k},s)$ is the operator of creation of a photon with momentum
${\bf k}$ and helicity $s$. We assume that the state amplitudes have 
non-overlapping supports
\be
\mbox{supp}\mbox{ }\psi_1({\bf k})\bigcap\mbox{supp}\mbox{ }\psi_2({\bf k})=\emptyset,
\quad \Omega_i({\bf k})=\mbox{supp}\mbox{ }\psi_i({\bf k}).
\en
Then the states $\ket{\psi_1}$ and $\ket{\psi_2}$ are orthogonal
\be
\BraKet{\psi_1}{\psi_2}=\int\psi_1^*({\bf k})\psi_2({\bf k})
\frac{d{\bf k}}{2k^0}=0.
\en
The measurement allowing to reliably distinguish between the two
orthogonal states $\ket{\psi_1}$ and $\ket{\psi_2}$ is given by the 
identity resolution
\be
{\cal P}_{\psi_1}+{\cal P}_{\psi_2}+{\cal P}_{\bot}=I,\quad
I=\int \ket{{\bf k},s}\bra{{\bf k},s}\frac{d{\bf k}}{2k^0},
\en
\bem
{\cal P}_{\psi_{1,2}}=
\left(\int_{\Omega_{1,2}({\bf k})} 
\psi_{1,2}({\bf k})\ket{{\bf k},s}
\frac{d{\bf k}}{\sqrt{2k^0}}\right)
\left(\int_{\Omega_{1,2}({\bf k})}
\bra{{\bf k}',s}\psi^{*}_{1,2}({\bf k}')
\frac{d{\bf k}'}{\sqrt{2k^{0'}}}\right),\quad
{\cal P}_{\bot}=I-{\cal P}_{\psi_1}-{\cal P}_{\psi_2}.
\enm
The probabilities of different outcomes for the input state 
$\ket{\psi_1}$ are given by the relations
\be
\mbox{Pr}_1\{ {\cal P}_{\psi_1} \}=\mbox{Tr}\{\ket{\psi_1}\bra{\psi_1}{\cal P}_{\psi_1} \}=
\left(\int_{\Omega_1({\bf k})} |\psi_1({\bf k})|^2\frac{d{\bf k}}{2k^0}\right)
\left(\int_{\Omega_1({\bf k}')} |\psi_1({\bf k}')|^2\frac{d{\bf k}'}{2k^{0'}}\right)
\equiv 1,
\en
\bem
\mbox{Pr}_{\psi_1}\{ {\cal P}_{\psi_2,\bot} \}=\mbox{Tr}\{\ket{\psi_1}\bra{\psi_1}
{\cal P}_{\psi_2,\bot} \}\equiv 0,
\enm
and similarly for the input state $\ket{\psi_2}$.

To illustrate the non-local nature of the measurement (13) more clearly,
let us consider an auxiliary measurement in the position representation which
transforms to the measurement allowing reliable distinguishing of the two states
when the involved spatial domain is extended to the entire space.

Consider first an auxiliary measurement which allows to reliably distinguish
between any two states with non-overlapping supports rather than a particular 
pair of states. This measurement is very similar to the measurement (13):
\be
{\cal P}_1+{\cal P}_2+{\cal P}_{\bot}=I,\quad
I=\int \ket{{\bf k},s}\bra{{\bf k},s}\frac{d{\bf k}}{2k^0},
\en
\bem
{\cal P}_{1,2}=\int_{\Omega_{1,2}({\bf k})} 
\ket{{\bf k},s}\bra{{\bf k},s}\frac{d{\bf k}}{2k^0},\quad
{\cal P}_{\bot}=I-{\cal P}_1-{\cal P}_{2}.
\enm
The probabilities of different outcomes for the input state 
$\ket{\psi_1}$ are 
\be
\mbox{Pr}_1\{ {\cal P}_1 \}=\mbox{Tr}\{\ket{\psi_1}\bra{\psi_1}
{\cal P}_1 \}=\int_{\Omega_1({\bf k})} |\psi_1({\bf k})|^2
\frac{d{\bf k}}{2k^0}\equiv 1,
\en
\bem
\mbox{Pr}_{1}\{ {\cal P}_{2,\bot} \}=\mbox{Tr}\{\ket{\psi_1}\bra{\psi_1}
{\cal P}_{2,\bot} \}\equiv 0,
\enm
and similarly for the input state $\ket{\psi_2}$.

According to the quantum-mechanical measurement theory, any measurement
is described by the positive identity resolution on the Hilbert space of 
the system states, or, to be more precise, by the positive operator-valued 
measure $M$ (on a certain measurable outcome space
${\cal U}$) satisfying the conditions [13]
\bem  
\mbox{1)}\quad M(\emptyset)=0,\quad M({\cal U})=I,\quad M({\cal U}_i)\ge 0, 
\enm 
\be 
\mbox{2)}\quad M({\cal U}_1)\le M({\cal U}_2),\quad 
{\cal U}_1\subseteq{\cal U}_2,\ \ \ \ \ \ \ \ \ \ \ \ \en 
\bem 
\mbox{3)}\quad M(\bigcup_i{\cal U}_i)=\Sigma_i M({\cal U}_i), \quad 
{\cal U}_i\bigcap {\cal U}_j=\emptyset.  
\enm
We shall choose a three-element discrete space whose points are labeled 
by the symbols $\{1,2,\bot\}$ as the space of outcomes: 
\be
M_{1,2}=\int_{ \Omega({\bf x}) } d{\bf x}
\left( \int_{\Omega_{1,2}({\bf k}) } e^{i\hat{ k}(\hat{x}-\hat{x}_0)}\ket{{\bf k},s}
\frac{d{\bf k}}{\sqrt{2k^0}}
\right)
\left( \int_{\Omega_{1,2}({\bf k}') }  e^{-i\hat{ k}'(\hat{x}-\hat{x}_0)}\bra{{\bf k}',s}
\frac{d{\bf k}'}{\sqrt{2k^{0'}}}
\right),
\en
\be
M_{\bot}=\int_{\cal X}\int_{\cal K}\int_{\cal K} d{\bf x}
\left(e^{i\hat{ k}(\hat{x}-\hat{x}_0)}\ket{{\bf k},s}
\frac{d{\bf k}}{\sqrt{2k^0}}
\right)
\left(e^{-i\hat{ k}'(\hat{x}-\hat{x}_0)}\bra{{\bf k}',s}
\frac{d{\bf k}'}{\sqrt{2k^{0'}}}
\right)- M_1-M_2,
\en
where
\bem
{\cal X}=\{{\bf x}:\quad {\bf x}\in (-\infty,\infty)\},
\enm
\bem
{\cal K}=\{{\bf k}:\quad  {\bf k}\in (-\infty,\infty)\}.
\enm
The positive operator-valued measure $M_i$ defines an identity resolution
on the space of outcomes $\{1,2,\bot\}$:
\be
M_1+M_2+M_{\bot}=
\int_{\cal X}\int_{\cal K}\int_{\cal K} d{\bf x}
\left(e^{i\hat{ k}(\hat{x}-\hat{x}_0)}\ket{{\bf k},s}
\frac{d{\bf k}}{\sqrt{2k^0}}
\right)
\left(e^{-i\hat{ k}'(\hat{x}-\hat{x}_0)}\bra{{\bf k}',s}
\frac{d{\bf k}'}{\sqrt{2k^{0'}}}
\right)=
\en
\bem
=\int_{\cal K} \ket{{\bf k},s}\bra{{\bf k},s}\frac{d{\bf k}}{2k^0}=I.
\enm

First of all, it is seen from (19,20) that when the spatial domain 
$\Omega({\bf x})$ is extended to the entire position space 
$\cal X$ the measurement transforms to the orthogonal identity resolution
(15) which allows to reliably distinguish between the states 
with non-overlapping supports.

For example, the probability of obtaining the outcome 1 for the input state
$\ket{\psi_1}$ is
\bem
\mbox{Pr}_1\{M_1\}=\mbox{Tr}\{\ket{\psi_1}\bra{\psi_1}M_1\}=
\enm
\be
\int_{\Omega({\bf x})}d{\bf x}
\left(
\int_{\Omega_{1}({\bf k}) }
e^{i\hat{ k}(\hat{x}-\hat{x}_0)}\psi^*_1({\bf k})\frac{d{\bf k}}{2k^0}
\right)
\left(
\int_{\Omega_{1}({\bf k}) }
e^{-i\hat{ k}'(\hat{x}-\hat{x}_0)}\psi_1({\bf k}')\frac{d{\bf k'}}{2k^{0'}}
\right)=
\en
\bem
=-(2\pi)^6\int_{\Omega({\bf x})}d{\bf x} \mid\psi_1(-i\frac{\partial}{\partial{\bf x}})\mid^2
D^-_0(\hat{x}-\hat{x}_0)D^+_0(\hat{x}_0-\hat{x}).
\enm
Since the amplitude $\psi_1({\bf k})$ has a finite support, i.e. it vanishes
outside the domain $\Omega_1({\bf k})$, the domain of integration over 
${\bf k},{\bf k'}$ in Eq.(21) can be extended to the entire space ${\cal K}$. 
We assume the amplitude 
$\psi_1({\bf k})$ to be a sufficiently smooth function to allow 
the substitution of the argument ${\bf k}$ by $i\partial/\partial{\bf x}$ 
followed by its extraction outside the integral sign over
${\bf k}$. Further, taking into account the definition of function 
$D^-_0(\hat{x})$ (8), one obtains the final expression. Remember
also that $D^-_0(\hat{x})=-D^+_0(\hat{-x})$.
The probability of obtaining outcome 2 for the input state
$\ket{\psi_1}$ is identically equal to zero (and similarly
for the outcome 2 with the input state $\ket{\psi_2}$):
\be
\mbox{Pr}_1\{M_2\}=\mbox{Tr}\{\ket{\psi_1}\bra{\psi_1}M_2\}=
\en
\bem
\int_{\Omega({\bf x})}d{\bf x}
\left(
\int_{\Omega_{2}({\bf k}) }
e^{i\hat{ k}(\hat{x}-\hat{x}_0)}\psi^*_1({\bf k})\frac{d{\bf k}}{2k^0}
\right)
\left(
\int_{\Omega_{2}({\bf k}) }
e^{-i\hat{ k}'(\hat{x}-\hat{x}_0)}\psi_1({\bf k}')\frac{d{\bf k'}}{2k^{0'}}
\right)
\equiv 0,
\enm
since the amplitudes $\psi_1({\bf k})$ and $\psi_2({\bf k})$ possess 
non-overlapping supports.

Finally, the probability of obtaining the outcome $\bot$ for the input 
state $\ket{\psi_1}$ (and, similarly, for $\ket{\psi_2}$) is
\bem
\mbox{Pr}_1\{M_{\bot}\}=\mbox{Tr}\{\ket{\psi_1}\bra{\psi_1}M_{\bot}\}=
\enm
\be
\int_{{\cal X}\backslash\Omega({\bf x})}d{\bf x}
\left(
\int_{\Omega_{1}({\bf k}) }
e^{i\hat{ k}(\hat{x}-\hat{x}_0)}\psi^*_1({\bf k})\frac{d{\bf k}}{2k^0}
\right)
\left(
\int_{\Omega_{1}({\bf k}) }
e^{-i\hat{ k}'(\hat{x}-\hat{x}_0)}\psi_1({\bf k}')\frac{d{\bf k'}}{2k^{0'}}
\right)=
\en
\bem
=-(2\pi)^6 \int_{{\cal X}\backslash\Omega({\bf x})}d{\bf x}
\mid\psi_1(-i\frac{\partial}{\partial{\bf x}})\mid^2
D^-_0(\hat{x}-\hat{x}_0)D^+_0(\hat{x}_0-\hat{x}).
\enm
The measurement (19--23) has a simple meaning. The equation (21) 
describes the probability of detection in the spatial domain
$\Omega({\bf x})$ of the state whose support lies in
$\Omega_1({\bf k})$ (and similarly for $\ket{\psi_2}$).  

Reliable detection of the states whose support lies in $\Omega_1({\bf k})$
requires the access to the entire spatial domain where the state is present;
it is seen from Eq.(21) that contribution to the probability is only
given by the causally related points 
($(\hat{x}-\hat{x}_0)^2=0$, $|{\bf x_0}-{\bf x}|=c|t_0-t|$) 
since the commutator functions $D_{0}^{\pm}(\hat{x}-\hat{x}_0)$ 
are only different from zero on the light cone. In spite of the fact that
the commutation functions in (21,22) both have a singularity on the light cone,
their product always exists as a distribution since the Fourier transforms
of $D^-$-functions have their supports in the front part of the light cone
(for details see e.g. [14]).

Equation (23) yields the probability of detection of a state with
the support in $\Omega_1({\bf k})$ (and similarly for $\ket{\psi_2}$)
in the rest part of the space ${\cal X}\backslash\Omega({\bf x})$ 
due to the ``tails'' of the state $\ket{\psi_1}$ which do not ``fit'' into
the spatial domain $\Omega({\bf x})$. When the accessible domain
$\Omega({\bf x})$ is extended to the extent that the entire
state $\ket{\psi_1}$ can ``fit'' into it, the measurement discussed
transforms to the measurement corresponding to the orthogonal projectors (15)
which allows to reliably (with unit probability) distinguish between the 
two states, the probability of detection of the ``tails'' $\bot$ tending 
to zero.

Eq.(21) is especially transparent for the state which is strongly localized 
in the position space. In that case the state amplitude in the momentum space
is strongly delocalized (in the limit 
$|\psi_1({\bf k})|^2)\rightarrow {\rm const}$ the support 
$\Omega_1({\bf k})\rightarrow{\cal K}$ and, accordingly, 
$|\psi (-i\partial/\partial {\bf x} )|^2$ does not depend on ${\bf x}$). 
Eq. (21) becomes
\be
\mbox{Pr}_1\{M_1\}=
(2\pi)^6\int_{\Omega({\bf x})}d{\bf x} \mid D^-_0(\hat{x}-\hat{x}_0)\mid^2.
\en
If we now remember that the function $-iD^-_0(\hat{x}-\hat{x}_0)$ 
describes the amplitude of the propagation of a one-particle field state
created at a point $\hat{x}_0$ to the point $\hat{x}$
\be
\bra{0} \psi^{-}_{1} (\hat{x}) \psi^{+}_{1} (\hat{x}_0) \ket{0}=
-i D^{-}_{0} (\hat{x}-\hat{x}_0), \quad x_{0} > x^{0}_{0},
\en
then Eq.(21) yields the probability
of detection of the one-particle field state with the support in
$\Omega_1({\bf k})\rightarrow{\cal K}$
in the spatial domain $\Omega({\bf x})$. It is seen from Eq.(24), 
that if the integration domain $\Omega({\bf x})$ does not include the points
where the one-particle state was created 
$\hat{x}_0$ the detection probability is zero.
For reliable detection of a strongly spatially localized state the 
spatial domain should be chosen arbitrarily small,
$\Omega({\bf x})\rightarrow 0$. The field can ``fill in'' this
domain arbitrarily fast (naturally, excluding the time-of-flight 
from the creation point  $\hat{x}_0$ to $\Omega({\bf x})$).

On the contrary, if the state is not strongly localized in space
(so that $|\psi_1({\bf k})|^2\neq const$ and, accordingly, 
$|\psi_1(-i\partial/\partial{\bf x})|^2\neq const$), the contributions
are given by all the points of the domain $\Omega({\bf x})$. 
The stronger the state is localized in the ${\bf k}$-space, 
the stronger it is delocalized in the position space and the larger
domain $\Omega({\bf x})$ is required for a reliable (with unit probability) 
detection of the state. The field cannot ``fill in'' that domain
in time less than $t\approx L(\Omega({\bf x}))/c$, where $L$ is the 
characteristic domain size.
              
It will be important for the bit commitment protocol that
the detection probability grows with time (as the field fills the domain
accessible to the measurement) and finally after the time $T$ elapses
(this time depends on the structure of the particular state) the states
become reliably distinguishable due to their orthogonality. During
the time interval $0<t<T$ the states are effectively 
non-orthogonal (non-distinguishable reliably). Choosing the supports
$\Omega_{1,2}({\bf k})$ of the states more and localized the time
interval $T$ can be made arbitrarily long.

Because of the symmetry between 0 and 1 it is convenient to choose 
the states in such a way that the domain $\Omega({\bf x})$ be the same
for them (the correct identification probabilities as functions of time will 
then be identical for 0 and 1). For the photons this can always be done by 
choosing, for example, a pair of narrow-band state with the same frequency
width and different central frequencies.

It should be emphasized that Eqs.(21--23) are statistical in nature.
If the domain accessible in the measurement procedure is small, or
equivalently, the state should be identified in a short time, the probability
of the correct answer is small, since most of the outcomes will occur in the 
channel $M_{\bot}$ independently of the income state. The probabilities of the
outcomes 1 and 2 are small due to the smallness of the domain accessible
to the detection procedure. However, firing of the detector in the channel 
$M_1$ or $M_2$ is sufficient to correctly identify the state, the problem 
being the low probability of the corresponding events. 

It is also easy to write down the measurement involving a finite spatial 
domain and allowing to reliably distinguish between a pair of
particular orthogonal states $\ket{\psi_1}$ and $\ket{\psi_2}$ 
(rather than the states with non-overlapping supports)
when the spatial domain $\Omega({\bf x})$ is extended to the entire 
position space:
\be
M_{1,2}=
\en
\bem
\left(\!\int_{ \Omega({\bf x}) }\!\! d{\bf x}\!\int_{\cal K}\!\int_{\cal K}
\!\!\psi_{1,2}({\bf k})e^{i(\hat{k}-\hat{k}')(\hat{x}-\hat{x}_0) }
|{\bf k}\rangle
\frac{ d{\bf k} d{\bf k}' }{ (2k^0 2k^{0'})^{1/4} }\!\right)\!\!
\left(\!\int_{\Omega({\bf x})}\!\! d{\bf x}\!\int_{\cal K}\!\int_{\cal K}
\!\!\langle{\bf k}|
\psi^{*}_{1,2}({\bf k})e^{-i(\hat{k}-\hat{k}')(\hat{x}-\hat{x}_0)}
\frac{ d{\bf k} d{\bf k'} }{ (2k^0 2k^{0'})^{1/4} }\!\right),
\enm
\be
M_{\bot}=I-M_1-M_2.
\en
When the accessible domain is extended to the entire space
($\Omega({\bf x})\rightarrow {\cal X}$) this measurements transforms to 
the orthogonal projectors given by Eq.(13). 

\section{Example of a one-dimensional measurement}

To get a qualitative picture and obtain more accurate estimates,
we shall consider a model one-dimensional situation since the 
three-dimensional case requires the specification of the geometry
of spatial domains. In addition, the experimental realization usually employ
the optical fiber (which is a one-dimensional system) as the quantum channel.
Similar model one-dimensional schemes are frequently used in quantum optics.

Suppose we have a pair of orthogonal single-photon packets
\be
\ket{\psi_{1,2}}=\int_{0}^{\infty}\psi_{1,2}(k)\ket{k}dk,\quad
\rho_{1,2}=\ket{\psi_{1,2}}\bra{\psi_{1,2}},\quad k>0,\quad
\BraKet{k}{k'}=\delta(k-k'),
\en
where $\ket{k}=a_k^+\ket{0}$ is the single-photon
monochromatic Fock state (we consider the particles moving
in only one direction). In addition, in the one-dimensional case,
$k^0=k$ ($c=1$ is the velocity of light).

The states have non-overlapping supports with the same bandwidth
\be
E_{1,2}=\mbox{supp}\mbox{ }\psi_{1,2}=\{k:\quad k\in(-\Delta/2+k_{1,2},k_{1,2}+\Delta/2)\}, 
\quad
E=\{k:\quad k\in[0,\infty)\}, 
\en 
\bem 
k_1-k_2\ge\Delta, \quad \int_{E_{1,2}}|\psi_{1,2}(k)|^2dk=1, 
\enm 
The states with the non-overlapping supports are orthogonal
\be
\BraKet{\psi_1}{\psi_2}=\int_{0}^{\infty}\psi^{*}_{1}(k)\psi_2(k)dk=0.
\en
The measurement allowing to reliably distinguish
between the states (28) is given by the identity resolution similar to Eq.(15)
\be
{\cal P}_1+{\cal P}_2+{\cal P}_{\bot}=I,\quad 
I=\int_{0}^{\infty}\ket{k}\bra{k}dk,\quad
{\cal P}_{1,2}=\int_{E_{1,2}}\ket{k}\bra{k}dk,\quad
{\cal P}_{\bot}=\int_{E\setminus(E_1\bigcup E_2)}\ket{k}\bra{k}dk.
\en
The identity resolution (31) does not involve the position and time parameter
indicating that it is implicitly assumed that the spatial domain accessible
to the measurement procedure is $x\in(-\infty,\infty)$.

In the one-dimensional case the measurement $M_i$ similar to (19,20) has the 
same form, the only difference being the replacement of the three-dimensional
domain $\Omega({\bf x})$ by the one-dimensional which for brevity will be 
denoted as ${\cal X}=\{ x: \quad x \in (-X,X) \}$.
The one-dimensional case is especially transparent, since the time parameter
and position appear in the combination
$x-ct$ which is actually related to the properties of the fundamental
solution of the one-dimensional wave equation which is known to have the form
${\cal E}(x,t)=\propto\theta(ct-|x|)$ unlike the three-dimensional case
(${\cal E}(x,t)\propto\theta(t)\delta(c^2t^2-|{\bf x}|^2)$) [15].
Therefore, in the one-dimensional case expansion of the spatial domain
$X$ accessible in the measurement can be effectively obtained 
(for fixed $X$) by increasing the accessible time interval $T$. 
Taking the above consideration in the account we shall below speak for
brevity that the measurements are performed in the accessible time 
window $(-T,T)$.

The probability of occurrence of the outcome 1 for the input density matrix
$\rho_1$ in the one-dimensional case is conveniently written 
(first performing integration over $x$) as
\be
\mbox{Pr}_1\{ M_1 \}=\Tr\{\rho_1 M_1\}=
\int_{E_1}\int_{E_1}\psi^*_1(k)\psi_1(k')
\frac{\sin{[(k-k')T]}}{(k-k')}
dk dk',
\quad T \equiv ct-|X|.
\en

Eq.(32) yields the probability of detection of the systems 
whose density matrix has the support in $E_1$ during the time window 
$(-T,T)$. 

If the time window during which the measurements are allowed is short
$(T\Delta\ll 1)$, Eq.(32) yields
\be
\mbox{Pr}_1\{\ M_1 \}\approx T\Delta\ll 1 \approx 0,
\en
i.e. the detection probability is proportional to the time window duration.
For a long time window $(T\Delta\gg 1)$ one has
\be
\mbox{Pr}_1\{\ M_1 \}\approx \frac{1}{\pi}
\int_{-T}^{T}\frac{\sin{\xi}}{\xi}d\xi=1-\cos{(T\Delta)}/T\Delta\approx 1,
\quad  T\Delta \gg 1 .
\en
If the input state is $\rho_2$, the probability of obtaining outcome 2
is zero independently of the time window duration $T$ (and similarly
for $\rho_1$ and outcome 2):
\be
\mbox{Pr}_2\{ M_1 \}=\Tr\{\rho_2 M_1\}=
\mbox{Pr}_1\{M_2\}=\Tr\{\rho_1 M_2\}=0.
\en
Eq.(32) can be rewritten in a more usual form as
\be
\mbox{Pr}_1\{M_1\}=\int_{-T}^{T}|\psi_1(\tau)|^2d\tau,\quad
\psi_1(\tau)=\int_{0}^{\infty}\psi_1(k)e^{-ik\tau}dk,
\en
which is consistent with the intuitive classical concepts.
If the function is localized in the frequency representation, it is
delocalized in the temporal representation, so that reliable signal
detection requires a long time window which can fully ``cover''
$\psi(\tau)$.

Here one reservation is in place. If we are given a classical state, i.e. 
a classical function $\psi(x)$ with the beforehand unknown support
in the $k$-representation ($\mbox{supp}\mbox{ }\psi(k)=\Delta$), 
the support can only be found if one finds the values of the function in
the spatial domain whose size is not less than $L_x \approx 1/\Delta$.  
On the other hand, if one has to distinguish between the two
classical functions $\psi_1(k)$ and $\psi_2(k)$ with non-overlapping
supports it is sufficient to know the function values in a single
point $x$ where $\psi_1(x)\neq\psi_2(x)$. Therefore, reliable distinguishing
between the two classical functions requires the knowledge of the function 
values in a single point where these values are different.

To reliably distinguish between the two orthogonal quantum states 
$\ket{\psi_1}$ and $\ket{\psi_2}$ with unit probability one requires the 
access to the entire spatial domain where these states are different from 
zero (we assume that both states are different from zero in the same domain).

For a short time window ($T\ll 1/\Delta$, where $\Delta$ is the bandwidth) 
the detection probability is small to the extent $T\Delta\ll 1$. 
If the detection occurs in channels 1 or 2, this is sufficient
to reliably distinguish between the states 
$\ket{\psi_1}$ and $\ket{\psi_2}$. However, for
a short time window dominating for both input states 
$\ket{\psi_1}$ and $\ket{\psi_2}$ will be the detection events in channel
$\bot$, i.e. the states will be indistinguishable with the probability 
close to 1 (effectively non-orthogonal).

The outcomes in channel $\bot$ will occur in the time window 
$(-\infty,\infty)$ due to the states whose supports do not lie in 
$E_1$ or $E_2$ and, in addition, in the time intervals 
$(-\infty,-T)$ and $(T,\infty)$ due to the ``tails'' of the states 
with the supports in $E_{1,2}$ which were not detected in the channels 
${1,2}$ in the time window $(-T,T)$.

The detection probability of $\rho_{1,2}$ in the channel $\bot$ is
\be
\mbox{Pr}_{1,2}\{ {\bot}\}=\int_{E_{1,2}}\int_{E_{1,2}}\left\{
\delta(k-k')-\frac{\sin{[(k-k')T]}}{(k-k')}\right\}
\psi^{*}_{1,2}(k)\psi_{1,2}(k')dk dk'.
\en
For $T\Delta\gg 1$  (long time window) we have
\be
\mbox{Pr}_{1,2}\{{\bot}\}\approx \cos{(T\Delta)/T\Delta}\rightarrow 0,
\en
while for $T\Delta\ll 1$  (short time window)
\be
\mbox{Pr}_{1,2}\{ {\bot} \}\approx \sin{(T\Delta)}/T\Delta
\rightarrow 1.
\en
Eqs.(32--39) actually mean that if one of the states 
$\rho_1$ or $\rho_2$ is given and one should identify which state
is actually given during the time interval $T$ then 
for $T\ll 1/\Delta$ the probability of the correct identification 
$p_+\approx T\Delta\ll 1$ (accordingly, the wrong answer probability is 
$p_-\approx 1-T\Delta\sim 1$), since the detection probability 
in the time interval $T$ is small. On the other hand if
the measurements are allowed to be performed in th long time 
window $(T\Delta\gg 1)$, the probability of the correct answer
$p_+\approx 1-1/T\Delta \approx 1$ so that the states are
reliably distinguishable.

For a short time window the sates are effectively non-orthogonal
(they cannot be distinguished reliably). The effective angle 
$\alpha$ between the states $\rho_1$ and $\rho_2$ in the time interval
$(-T,T)$ is small
\be
\BraKet{\psi_1}{\psi_2}\approx \cos{\alpha}\approx 1-T\Delta,\quad
\alpha\approx T\Delta\ll 1.
\en

\section{Relativistic quantum bit commitment protocol}

Let us now describe the bit commitment protocol. The two parties agree in 
advance on the states $\ket{\psi_1}$ and $\ket{\psi_2}$ corresponding to 0 
and 1 as well as on the number $N$ of quantum systems
sent by user A to B. It is always possible to choose a pair of orthogonal states so that 
their spatial extent strongly exceeds the channel length and their 
fall-off at the infinity insures almost reliable identification after the 
time $T$ elapses. In the case of photons one can choose two states
with sufficiently narrow energy spectra so that their effective extent
$L\approx c/\Delta\omega$ ($\Delta\omega$ is the energy spectrum width) 
substantially exceeds the communication channel length $L_{ch}$, 
In that case the channel length can effectively assumed to be zero.
Formally that means that the two parties A and B can only control
the neighbourhoods of the points 
$x_A$ and $x_B$ and do not control the space outside these neighbourhoods.
In other words, outside the neighbourhoods of $x_A$ and 
$x_B$ both parties are allowed to do anything which 
does not contradict the laws of relativistic quantum mechanics.

The user A chooses his bit $a$ (0 or 1), which is the parity bit
of $N$ quantum states ($a=a_1\oplus a_2\oplus\ldots\oplus a_N$).
Then at time $t=0$ the user A sends to B all $N$ states simultaneously. 
To be more precise, user A turns on $N$ sources producing
the states which are allowed to propagate into the communication channel
as long as they are being formed. 
Since the states $\ket{\psi_1}$ and $\ket{\psi_2}$ are generally non-stationary,
fixing the time moment $t=0$ allows user B to  
``tune''  the measurement (13) by choosing appropriate phase shifts for 
any spatial domain reached by the field as it is propagating in the communication 
channel. User B performs measurements (13) separately on each state.
Because of the orthogonality, the states become distinguishable by the time 
$T$ when they are first fully accessible.
In the time interval $0<t<T$ the states are not fully accessible
(effectively non-orthogonal) and therefore are not reliably distinguishable.
The probability of correct identification of the states is an increasing
function of time
($p(t=0)=0$ and $p(t=T)=1$), the specific 
form of $p(t)$ depending on the structure of the particular states
used and being irrelevant for our analysis.
For individual measurements the probability of the
correct identification of the parity bit
$P_+(t)=p^N(t)\ll 1$ as long as $p(t)\ll 1$. 
For collective measurements, when the measurements are performed over all
$N$ states as a whole the probability of the correct parity bit identification
is $P_{+}^{collect}(t)=\sqrt{p^N(t)}\ll 1$ [16] and can also be made
arbitrarily small by appropriate choice of the states used. The states 
can always be chosen in such a way that for an arbitrarily small quantity
$\delta\ll 1$ specified beforehand and arbitrarily large
time $t_c$ ($t_c$ is the time of secure storage of the secret bit chosen 
by user A) the probability $P_{+}^{collect}(t)$ of the correct
identification of the parity bit during time $t<t_c$ is arbitrarily
small. This can be achieved by increasing the effective extent
of the states (reducing the spectrum width).

The above discussion concerns the case where the user B obtains the information
on the states only from the quantum-mechanical measurements. The mentioned
probabilities $p(t)$ actually represent the probabilities of the state
detection (firing of a classical apparatus) which for times $t<T$ is less
than 1 because the entire state is not accessible for the measurement.
However, if the detection did take place, the state can be identified
by the obtained measurement outcome. Therefore, as long as the probability of
the detection itself $p_i(t)<1/2$, the user B can even perform no 
measurements simply guessing the states sent in each communication channel.
However, for the times when $p_i(t)>1/2$ 
the measurements provide more information than simple guessing without 
performing any measurements. For $t\rightarrow T$ the measurements
provide almost reliable information on the states.

Therefor, the probability $P^{store}(t)$ 
of secure storage of the secrete bit chosen by user A is a decreasing
function of time ($P^{store}(t=0)=1$ and $P^{store}(t=T)=0$). 

After the quantum part of the protocol is completed at $t>T$, when
user B already has the full access to the states and can reliably identify
them, user A discloses through a classical channel which states 
he sent through each of $N$ quantum channels. Any inconsistency between
the classical information and the measurement result in at least one channel
aborts the protocol. The necessity of conveying the classical information
by user A after the quantum-mechanical measurements and the individual
reliable distinguishability (orthogonality) of the states eliminates
the possibility of cheating. For example, user A cannot send the mixed
states like $\rho=\ket{\psi_1}\bra{\psi_1}+\ket{\psi_2}\bra{\psi_2}$, 
since for large $N$ this would result in the discrepancy between the
classical information and the outcomes of quantum measurements. Neither can he
send any other states instead of $\ket{\psi_1}$ and $\ket{\psi_2}$ because
of their orthogonality, i.e. reliable distinguishability. Any different
states would yield for large $N$ wrong outcomes if the measurement
${\cal P}_{1,2,\bot}$ is used. The constraints on the maximum field
state propagation velocity does not allow the user A to delay
sending his states to user B since the measurement (13) is ``tuned''
to precisely the states $\ket{\psi_1}$ and $\ket{\psi_2}$ so that the 
delayed states would produce wrong outcomes. Formally, the delay
can be described as a translation in the space-time which results
in an additional phase factor $e^{i\hat{k}\hat{x}_0}$ under the integral 
sign in (10). The ``shifted'' state, e.g. $\ket{\psi_1}$, will not
yield the outcome 1 with the unit probability.
 
In the non-relativistic case the user A can employ the EPR pair
which allows him to delay his choice of the 
secret bit until the measurement performed by user B [5,6].
In the relativistic case the EPR-attack does not work because of
the constraint on the maximum speed of the field state propagation.
In spite of the fact that the EPR correlations are also preserved in 
the relativistic case for the measurements performed on the entangled 
two-particle states of the field at the points separated by the 
space-like interval [17], the field cannot propagate from the EPR-source
faster than light. Therefore, in contrast to the non-relativistic
quantum mechanics, employment by user A of an EPR-pair to delay
his choice until the second stage of the protocol does not work.

It should be noted once again that it is important for the protocol that
the orthogonal states have quantum nature, so that their reliable 
distinguishability (with unit probability) requires a finite spatial domain.
Two classical states (functions) can be reliably distinguished by
their values at a single suitably chosen point.

The time interval $T$ required for a reliable distinguishability of the states,
i.e. the time of secure storage of the secrete bit, is determined by the 
spectrum bandwidth $\Delta |{\bf k}|\approx \Delta\omega/c$ of the photons
used in the protocol and is estimated as $T\approx 1/\Delta\omega$. 
Note that although there are no any fundamental constraints
on making the photon bandwidth arbitrarily small, this problem
is technically very difficult.

\section{Conclusions}

We conclude with the following remark. The possibility of the state
identification with the unit probability during time interval $T$
depends on the existence for a particular type of particles of the 
states with finite spatial support. For the photons only the states with
the exponentially localized energy and detection rate are known at present 
[18]. The latter formally means that the correct identification
with unit probability is only possible for an infinite time interval.
This consideration is not too restrictive fro the protocol since the 
time interval can be chosen to be sufficiently long to ensure
the distinguishability probability exponentially close to unit
for two orthogonal states.

The experimental realization is rather simple (at least in principle).
It is sufficient to have $N$ broadband sources. The term ``broad band'' here 
means that when turned on these sources produce the states localized in space
and time, that is at the time moment $t_0$ one can prepare the signals
with wide frequency spectrum. Before being allowed in the communication
channel, these states should pass the narrow-band filters and attenuators 
to reach the one-photon level. Passage through the narrow-band filter
requires either long time or large space. The detection is performed
by the narrow-band detectors.

This work was supported by the Russian Foundation for Basic Research
(project No 99-02-18127), grant No 02.04.5.2.40.T.50 
within the framework of the Program ``Advanced devices and technologies 
in micro- and nanoelectronics'' and by the Program ``Physics of quantum 
and wave processes'' (subprogram ``Quantum computing'').

\end{document}